\documentstyle{elsart} 
\begin{document}

\begin{frontmatter} \title{ Stability of Spatio-Temporal Structures in
a Lattice Model of Pulse-Coupled Oscillators }

\author[Barna]{A.  D\'{\i}az-Guilera}, \author[Tarragona]{A.  Arenas},
\author[Barna]{A.  Corral}, and \author[Barna]{C.  J.  P\'{e}rez}

\address[Barna]{ Departament de F\'{\i}sica Fonamental, Facultat de
F\'{\i}sica,\\ Universitat de Barcelona, Diagonal 647, E-08028
Barcelona, Spain} \address[Tarragona]{Departament d'Enginyeria
Inform\'atica, Universitat Rovira i Virgili, \\ Carretera Salou s/n,
E-43006 Tarragona, Spain}


\begin{abstract}
We analyze the collective behavior of a lattice model
of pulse-coupled oscillators.  By studying the intrinsic dynamics of
each member of the population and their mutual interactions we observe
the emergence of either spatio-temporal structures or synchronized
regimes.
We perform a linear stability analysis of these structures.  
\end{abstract} 
\end{frontmatter}
%
%
\section{Introduction}
The collective behavior of large assemblies of pulse-coupled
oscillators has been investigated quite often in the last years.  Many
physical and biological systems can be described in terms of
populations of units that evolve in time according to a certain
intrinsic dynamics and interact when they reach a threshold value
\cite{HSS}.  Although it was known long time ago that the members of
these systems tend to have a synchronous temporal activity, a rigorous
treatment of the problem has been considered only in the last decade
\cite{Winfree,Ermen,MS}.  Up to now, the most important efforts have
been focused on systems with long-range interactions because in this
case analytical results can be derived by applying a mean-field
formalism.  Relevant is the work by Mirollo and Strogatz (MS)
\cite{MS} who discovered under which conditions mutual synchronization
emerges as the stationary configuration of the population.  Later on,
the study has been generalized to other situations
\cite{Kura,Chiu,CA,Abbpre,EPG,PRL2}.

When the oscillators form a finite dimensional lattice where only
short-range interactions are allowed, the spectrum of behaviors is
broader.  For example, under certain conditions lattice models of
pulse coupled-oscillators display self-organized criticality (SOC)
\cite{tesikim,PRL,MT,Bottani,HopHerz,Review}.  In such case the system
self-organizes, due to its own dynamics, into a critical state with no
characteristic time or length scales and provide some hints about the
wide ocurrence of $1/f$ noise in nature \cite{BTW}.  In other cases,
phase locking, synchronization, or more complex spatio-temporal
structures are developed.

On the other hand, some efforts have been devoted recently to
analyze the stability of spatio-temporal periodic structures in
coupled map lattices \cite{chinchon}.  These systems are easier to
deal with, both analytically and numerically, than lattices of coupled
nonlinear oscillators although they represent a less realistic view
of coupled systems that appear quite often in nature.

It is precisely the purpose of the present work to study the stability
of certain structures, that have been obtained recently in computer
simulations in a lattice of pulse-coupled oscillators.  Keeping this
goal in mind the paper is organized as follows.  
In the next section
we introduce the model and explain several equivalent ways to
describe the time evolution of the system. These equivalent 
descriptions
are related through well defined transformations,
which allow to study the model
in terms of the most suitable dynamic variables.
In Sect. 3, we calculate
analytically the fixed points of some of the structures observed in
simulations in one- and two-dimensional lattices,
and in Sect. 4 we analyze
the stability of these fixed points.  Finally, Sect. 5 is devoted to
the conclusions of this work and its future perspectives.
\section{The model}
Let us consider a system of coupled oscillators described each one by
a state variable $E_i$, which we can identify with a voltagelike
magnitude when dealing with biological oscillators. We will assume
that all the oscillators are identical and evolve in time according
to the following dynamics
\begin{equation}
\frac{dE_i}{dt}=f(E_i)+ \bar \varepsilon (E_i) \sum_{j}\delta
(t-t_j), \label{edete}
\end{equation} 
plus the reset condition for
$E_i\geq 1$. This means that when the $i$-th oscillator fires, it 
is reset ($E_i\geq 1\rightarrow 0$). The first terms of the r.h.s.
is the driving rate that throughout the paper will be considered
a positive quantity, $f(E) >0$. The second term accounts for the
coupling, given in terms of 
a nonlinear coupling function $\bar \varepsilon (E)$ that we will
assume either excitatory for all $E$, $\bar \varepsilon (E)>0$, or
inhibitory, $\bar \varepsilon (E)<0$, $\forall E$ (except at
the reset point $E=0$, where it could be $\bar \varepsilon (0)=0$).

When some of its neighbors (labeled by index $j$) fire at time 
$t_j$, the $i$-th oscillator suffers an
instantaneous perturbation of its state given by the coupling
function $\bar \varepsilon (E_i)$. This implies the existence of two
time scales, a slow one for the driving and a fast one for the
coupling. In this description both time scales appear in the same
equation. 
This represents a quite general evolution equation, because
not only the driving rate $f(E)$ but also the coupling $\bar
\varepsilon (E)$ are functions of the state $E$.

We could also consider another two equivalent descriptions of the same
model, both related to (\ref {edete}) by simple changes of variables.
The first equivalence is obtained by applying the following
nonlinear transformation 
\begin{equation} 
y(E)=\varepsilon
_0\int_0^E\frac{dE'}{\bar \varepsilon (E')} \label{ydee},
\end{equation} 
used before in different contexts by several authors
\cite{Abbpre,PRE}, where $\varepsilon _0$ is defined to ensure that
$y(E=1)=1$.  By substituting into (\ref{edete}) we get
\begin{equation} 
\frac{dy_i}{dt}= g(y_i)+\varepsilon
_0\sum_{j}\delta (t-t_j).  \label{ydete} 
\end{equation} 
Now, the
evolution of the system is described in terms of a new variable $y$
for which the coupling is constant and whose driving rate is given by
\begin{equation}
g(y_i)=\varepsilon_0\frac{f(E_i)}{\bar \varepsilon(E_i)}.  
\end{equation}
A case of
particular interest is that of a zero advance at the reset point
($\bar \varepsilon (E=0)=0$).  In such case, this transformation is
well defined if the coupling is constant ($\varepsilon _0$) $\forall
y\ne 0 $ except for $y=0$ where is exactly zero.  This condition plays
the role of a refractory time in the fast time scale, which provokes
that all the units which have fired have zero phase at the end of the
interactive process.

     The second equivalent description can be derived by writing
(\ref{edete}) as a function not of the state $E_i$ but of the phase of
each oscillator $\phi _i$, which evolves linearly in time except when
it receives a pulse from its neighbors.  In this
description, 
the driving and the
coupling are integrated in a 'phase response curve' (PRC), function
that gives the phase advance of the oscillator that receives the pulse
\cite{KI,Torras},
\begin{equation} 
\frac{d\phi _i}{dt}=1+\bar \Delta
(\phi_i)\sum_{j}\delta (t-t_j). \label{fidete} 
\end{equation}
This equation can be obtained in a straightforward manner by
applying the following transformation in (\ref{edete})
\begin{equation} 
\phi (E)=\int_0^E\frac{dE^{\prime }}{f(E^{\prime})}.
\label{fidee} 
\end{equation}
The fact that
$\phi(E=1)=1$ is guaranteed 
if the intrinsic period of the oscillators is equal to one.
According to the definitions the function $\bar \Delta $ is given by 
\begin{equation}
\bar\Delta(\phi_i) =\frac{\bar\varepsilon (E_i)}{f(E_i)}.  
\end{equation}
All three descriptions are physically equivalent, but depending on the
theoretical framework it is convenient to deal with one or
another for specific purposes.
In this paper we will use the description in terms of the phase.
Finally, it is worth noting that one has to be very careful when
handling with the two coexisting time scales in the system. Since they
do not overlap the situation is equivalent 
to stop the driving when an oscillator fires. Then, 
the resulting PRC $\Delta_n (\phi)$ must be related with
the function $\bar \Delta (\phi)$ by
\begin{equation} 
\int_\phi ^{\phi +\Delta_n (\phi)} \frac{d\phi
^{\prime }}{\bar \Delta (\phi ^{\prime
})}=\sum_{j}\int_{t^{-}_j}^{t^{+}_j}\delta (t-t_j) dt = n,
\label{deltavar} 
\end{equation}
where we obtain a different PRC depending on the number of firings $n$
that the oscillator receives, which will be bounded between one and
the number of neighbors.  This ensures that when $n$ neighbors fire
simultaneously, the $i$-th oscillator modifies its phase as 
\begin{equation} 
\phi_i \rightarrow \phi_i + \Delta_n (\phi_i).
\label{coupling.prc} 
\end{equation}   
otherwise it evolves as $d\phi_i/dt= 1$. In general, to go from 
$\bar \Delta (\phi)$ to $\Delta_n
(\phi )$ is straightforward, but the inverse implies to solve an
integral equation that in general can only be done
numerically.
\section{Fixed points}
     First of all, we will show that a lattice model of pulse-coupled
nonlinear oscillators evolving in time following 
(\ref{coupling.prc}) with
nearest-neighbor interaction and periodic boundary conditions
has some periodic spatio-temporal solutions which correspond
to fixed points in a return map description.
With this goal in mind, we will
study initially the behavior of a system of two oscillators and after
this we will generalize the analysis to one- and two-dimensional
lattice models.  Hereafter we will consider always refractory time
in the model, i.e., $ \Delta_n(\phi=0)=0$.
\subsection{Two oscillators}
     To start the discussion we will consider a system of two 
oscillators. It is simple enough to allow a complete description of 
the dynamical 
evolution of the model and, additionally, to illustrate the main 
ideas that we want to develop later on. We
present here a detailed evolution of the phase of each oscillator for 
half cycle. Each row corresponds to the phase of each one.
\begin{equation}
\begin{array}{ccccc}
1&\longrightarrow&0&\longrightarrow&1-\phi-\Delta_1(\phi)\\
&\mbox{firing}&&\mbox{driving}&\\
\phi &\longrightarrow& \phi + \Delta_1(\phi)&\longrightarrow&1
\end{array}
\label{returnmap}
\end{equation}
The fixed points of this transformation are the solutions
of $\phi ^{*}=1-\phi ^{*}-\Delta_1 (\phi ^{*})$. If  
$\Delta_1(\phi)$ is
a continuous function bounded between -1 and 1, there exists
at least one fixed point.  However an important
question arises about the uniqueness or multiplicity of those fixed
points.  It is easy to know under which sufficient conditions the
fixed point is unique by simple geometrical arguments,
which is $\Delta^{\prime}_1(\phi)>-2$, $\forall \phi$.  
It can be proved,
with our hypothesis of $f(E)>0$ and $\bar
\varepsilon(E) \ge 0$ or $\bar \varepsilon(E) \le 0$, $\forall E$,
that $\Delta^{\prime}_n(\phi)>-1$, 
$ \forall \phi$, verifying thus the uniqueness
of the fixed point.
To clarify this point, 
let us consider a particular example that allows
to calculate the location of the fixed point,  
the Peskin's model. It was proposed to
analyze the collective behavior of an assembly of cardiac pacemaker
cells \cite{Peskin}.  In this model the intrinsic time evolution 
of each member of
the population is given by
\begin{equation} 
f(E)=\gamma \left( K-E\right) , \label{peskin}
\end{equation} 
where $\gamma $ gives the slope of the driving rate and
$K=\left( 1-e^{-\gamma }\right) ^{-1}$.  The coupling function is for
this simple case a constant $\bar\varepsilon (E) = \varepsilon _0$.
Then the function $\bar \Delta$ is 
\begin{equation} \bar \Delta
(\phi )=\frac{\varepsilon _0}{\gamma (K-E)}=\frac{\varepsilon
_0}{\gamma K}e^{\gamma \phi }, \label{delvarpes} 
\end{equation}
where for the last equality it is necessary to know the relation
$E(\phi)$, substituting Eq. (\ref{peskin}) into Eq. (\ref{fidee}).  
Now
it is easy to find the PRC according to (\ref{deltavar}), obtaining
\begin{equation} 
\Delta_n (\phi )=-\frac 1\gamma \ln \left(1-\frac{n
\varepsilon _0}Ke^{\gamma \phi }\right), \label{delpes} 
\end{equation} 
It is straightforward to see from the derivative of $\Delta_n (\phi)$
that, for a given $\varepsilon_0>0$, the PRC is an increasing
function when $\gamma > 0$, whereas for $\gamma
<0$ the PRC decreases monotonously.  This
behavior is the opposite when an inhibitory coupling is taken
into account.
For two oscillators
the fixed point is unique and satisfies $\Delta_1
(\phi^{*})=1-2\phi^{*} $, corresponding to 
\begin{equation}
\phi^{*}=\frac{1}{2}-\frac{1}{\gamma} \sinh ^{-1} \left( \varepsilon_0
\sinh \left(\frac{\gamma} {2}\right)\right). \label{fxpes}
\end{equation}
In general for the two oscillators case we can study the global
stability of the system \cite{MS}.  To show this, we look at the result 
obtained
after half a cycle assuming identical oscillators.  Then for $\Delta_1
^{\prime}(\phi)>0$, any perturbation $ \delta $ 
from the fixed point evolves according to (\ref{returnmap}) as 
\[
\phi _{0}= \phi ^{*}+\delta \longrightarrow 1-\phi
^{*}-\delta - \underbrace{\Delta_1 (\phi ^{*}+\delta )}
_{>\Delta_1 (\phi
^{*})=1-2\phi ^{*}}
<\phi ^{*}-\delta 
\]
and 
\[
\phi _0= \phi ^{*}-\delta \longrightarrow >\phi
^{*}+\delta  .
\]
This behavior shows that the fixed point is a repeller of the
dynamics, when the PRC is an increasing function,
but due to the periodicity of the variable $\phi $ this
fact provokes synchronization between oscillators, i.e.,
oscillators are repelled from the fixed point and they go to 
phase one or zero which are the same cyclic point. 
The two oscillators approach each other after each
cycle until they are close enough so that the firing of one of 
the oscillators makes the phase of the other one to reach the threshold.
At this time, due to the refractory time, both oscillators
fire in unison and this situation persists forever, which corresponds
to the synchronized state.
In the same way, if $\Delta_1 ^{\prime}(\phi)<0$ we have:
\begin{eqnarray} 
\phi _{0} &=&\phi ^{*}+\delta \longrightarrow \qquad
>\phi ^{*}-\delta \nonumber \\ \phi _{0} &=&\phi ^{*}-\delta
\longrightarrow \qquad <\phi ^{*}+\delta \label{fis} 
\nonumber
\end{eqnarray}
Both equations need another limit 
\begin{eqnarray} 
\phi ^{*}+\delta
&\rightarrow& 1-\phi
^{*}-\delta-\Delta_1(\phi^{*}+\delta)<\phi^{*}+\delta \nonumber \\
\phi ^{*}-\delta &\rightarrow& 1-\phi
^{*}+\delta-\Delta_1(\phi^{*}-\delta)>\phi^{*}-\delta \label{fis2}
\nonumber
\end{eqnarray} 
where we have use that $\Delta_1(\phi)>1-2\phi$ if
$\phi>\phi^*$ and $\Delta_1(\phi)<1-2\phi$ if $\phi<\phi^*$.  
Now, for a decreasing PRC, the
fixed point is an attractor of the dynamics.  This attractor
corresponds to a phase-locked state that maintains the difference of
phases between both oscillators.
Note that we have assumed that the initial conditions are chosen in 
such a way that in the first interaction the oscillators do not
synchronize. This kind of synchronization is trivial and does not
depend on the slope of the PRC, only on its strength.

We have to mention, to clarify this point, that a PRC is always
bounded by $\Delta_n(\phi)<1-\phi$ when it is excitatory and by
$\Delta_n(\phi)>-\phi$ when inhibitory, in order to avoid phases
larger than $1$ or smaller than $0$.  However, these limits
correspond to trivial synchronization of the oscillators, and will
not be taken into account here.  Therefore, when we say increasing or
decreasing PRC we refer to the domain in which synchronization is not
trivial.
\subsection{One-dimensional lattices}
We have seen in the previous paragraphs that depending on the slope of
the PRC two oscillators tend either to synchronize or to keep a phase
difference between them.  When generalizing this result to a periodic
one-dimensional model of coupled oscillators with nearest-neighbor
interaction, we can imagine the same tendency.
On the one
hand the strength of the coupling makes two neighboring oscillators
with close phase values to synchronize.  On the other hand,
when they are not close enough, this
coupling can make the phases to approach each other 
or to separate, depending on
whether the slope of the PRC is positive or negative, respectively.
We have observed in computer simulations, starting from a
random distribution of phases, that the population reaches,
after a long transient and depending on the parameters of the system,
well defined spatial structures, with concrete values of the 
phase differences. We have checked that these structures are 
limit cycles for the dynamical evolution of the lattice. 
When building up the return map, by looking to the system at the time
that a given oscillator reaches the threshold, we obtain fixed points.

The most simple fixed point is that one in which all the oscillators
have exactly the same phase.  Thus all of them reach the threshold
simultaneously.  Having assumed that $\Delta_n (\phi =0)=0$, i.e.
the existence of a refractory time, all the oscillators will be at
zero phase after the collective firing
and then a new cycle starts again.  This corresponds to
global synchronization of the population.

There are other fixed points characterized by the fact that the
population is split into several subpopulations with the same number
of elements.  Each subpopulation is composed by oscillators with the
same phase, which is different for different subpopulations.
Moreover, they form periodic spatial structures and have been 
discussed 
in the context of coupled 
map lattices \cite{chinchon}.  A general case is
plotted in Fig.~1,
where the subscripts stand for the different subpopulations.
\begin{figure} 
\unitlength=1.00mm
\linethickness{0.4pt}
\begin{picture}(135.00,25.00)
\put(10.00,10.00){\framebox(15.00,15.00)[cc]{$\phi_1$}}
\put(25.00,10.00){\framebox(15.00,15.00)[cc]{$\phi_2$}}
\put(45.00,18.00){\makebox(0,0)[cc]{........}}
\put(50.00,10.00){\framebox(15.00,15.00)[cc]{$\phi_n$}}
\put(80.00,10.00){\framebox(15.00,15.00)[cc]{$\phi_1$}}
\put(95.00,10.00){\framebox(15.00,15.00)[cc]{$\phi_2$}}
\put(115.00,18.00){\makebox(0,0)[cc]{........}}
\put(120.00,10.00){\framebox(15.00,15.00)[cc]{$\phi_n$}}
\put(110.00,25.00){\line(1,0){10.00}}
\put(120.00,10.00){\line(-1,0){10.00}}
\put(50.00,10.00){\line(-1,0){10.00}}
\put(40.00,25.00){\line(1,0){10.00}}
\put(73.00,18.00){\makebox(0,0)[cc]{........}}
\put(65.00,10.00){\line(1,0){15.00}}
\put(80.00,25.00){\line(-1,0){15.00}}
\end{picture}
\caption{General 1d spatially
periodic structure.
Each $\phi_l$ corresponds to a different value of the phase.} 
\end{figure}
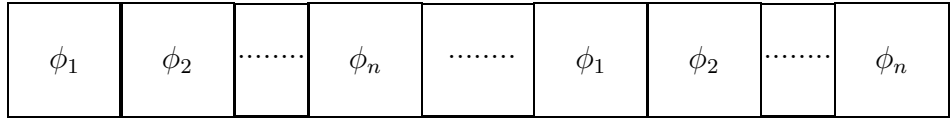
Among these structures we will focus on the simplest ones, those
having only two different phases, plotted in Fig.~2.  We
have introduced a notation to denote the number of synchronized ($s$)
or phase-locked ($a$) neighbors of a given oscillator that will
also be used for the 2d models.
\begin{figure}
\vspace{3.5cm} 
\caption{1d 2-phase fixed point
structures:  Top) Chessboard-like ($2a$); 
bottom) Domino-like ($1a1s$).}
\end{figure}
Without loss of generality we can assume that white sites in Fig.~2
have phase equal to one whereas black sites have a phase $\phi $.  
In these
cases it is very simple to compute the value of $\phi ^{*}$ that
correspond to the fixed points.  The proof runs parallel to that for
two oscillators. For the chessboard, if each row stands for a subpopulation

\begin{equation}
\begin{array}{ccccc}
1&\longrightarrow&0&\longrightarrow&1-\phi-\Delta_2(\phi)\\
&\mbox{firing}&&\mbox{driving}&\\
\phi &\longrightarrow& \phi + \Delta_2(\phi)&\longrightarrow&1
\\
\end{array}
\end{equation}
and the same for the domino writing $\Delta_1(\phi)$ instead of
$\Delta_2(\phi)$. Then the fixed points correspond to
\begin {eqnarray}
\nonumber
\mbox{chessboard: } \phi ^{*}=1- \phi ^{*}-\Delta_2 (\phi ^{*}) =1- \phi
^{*}-\Delta_2 ^{*}
\\
\mbox{domino: } \phi ^{*}=1- \phi ^{*}-\Delta_1 (\phi ^{*}) =1- \phi
^{*}-\Delta_1 ^{*}
\nonumber
\end{eqnarray}
In the next section we will analyze in detail the linear stability of
the chessboard structure ($2a$) in 1d and how to generalize this
result to other structures.  
\subsection{Two-dimensional lattices}
     In two-dimensional lattices, with periodic boundary conditions
and nearest-neighbor interactions,
we have also observed in computer
simulations the existence of well-defined patterns that are
fixed points for the discrete dynamics of the population.
Some of
the two-phases structures are plotted in Fig. 3,
together with the unique one-phase structure, the synchronized state. 
The phases of
the fixed points are very easily computed using the transformations
for the fixed point explained in the one-dimensional case.  
In general we are going to have
for the different possibilities between phase-locked ($a$) or
synchronized ($s$) neighbors in a 2d square lattice, the following
fixed point equations:  
\begin{eqnarray} 4a\qquad \phi ^{*} &=&1-\phi
^{*}-\Delta_4 ^{*} \nonumber \\ 3a1s\qquad \phi ^{*} &=&1-\phi
^{*}-\Delta_3 ^{*} \nonumber \\ 2a2s\qquad \phi ^{*} &=&1-\phi
^{*}-\Delta_2 ^{*}\nonumber \\ 1a3s\qquad \phi ^{*} &=&1-\phi
^{*}-\Delta_1 ^{*} \label{2dFP} \\ &&\qquad \nonumber 
\end{eqnarray}
\begin{figure} 
\vspace{22cm}
\caption{2d 2-phase fixed point structures and the synchronized
state ($4s$).}
\end{figure}
In our computer simulations we have observed that these
structures have different basins of attraction depending on
the values of the parameters as well as on the size of the lattice.
Nevertheless, the chessboard ($4a$) seems to be the most stable in
two senses: it has the largest basin of attraction and it is the
most robust to different sources of dynamical noise. In the next
section we will briefly discuss the linear stability of these 
patterns.
\section{Stability of the periodic structures}
     Now we will analyze the stability of the fixed points discussed
in the previous section.  We will discuss in detail the 1d
chessboard, and later on we will explain how to generalize this to
other structures.
\subsection{Linear stability analysis of the one-dimensional
chessboard}
To perform the stability analysis in the one-dimensional model we are
going to deal with the most simple configuration of two phases (black
and white) slightly perturbed at random from its fixed point
$(...,1,\phi ^*,1,\phi^*,...)$. 
Suppose without loss of generality
we have the subpopulation of white sites 
distributed slightly below phase one,
and the subpopulation of black sites
with phases around $\phi^*$, that is
\begin{eqnarray} 
&\mbox{White sites (W)} 
\left\{ \begin{array}{ll} 1 &
\hspace{3em}\mbox{first}\\ 1-\delta _1 & \hspace{3em}\mbox{last}
\end{array} \right.  
\nonumber \\ 
\end{eqnarray}
Where the top row stands for the first oscillator that will reach
the threshold and the bottom row for the last one, after
a first driving of amount
$\delta_1$.
On the other hand we have for black sites
\begin{eqnarray} 
&\mbox{Black sites (B)} \left\{
\begin{array}{ll} \phi^{*}+\delta_2 & \hspace{3em}\mbox{first}\\
\phi^{*}+\delta_3 & \hspace{3em}\mbox{last} \end{array} \right.
\label{iniconf} 
\end{eqnarray}
where the two rows have the same meaning as for the white sites.
Nevertheless, this does not mean that these oscillators have the
largest and smallest phases, in contrast to the white oscillators.
Since during the driving $\delta_1$ the black sites will receive 
pulses from their white neighbors at unknown phases, the order 
is not maintained and the black oscillators can overtake
each other.
Then the two oscillators shown in (\ref{iniconf}) are those that
after receiving the two pulses have the maximum and minimum phases,
respectively.
  
  As we have stated before, the first step is to wait for a driving
$\delta_1$; at this point all the white sites have fired and the 
new configuration becomes,
if in order to simplify
we call $\Delta$ to $\Delta_1$:

$
\mbox{W}\left\{ \begin{array}{l} \delta _1 \\ 0
\end{array} \right.  \label{wb} \\ 
\\
\mbox{B}\left\{ 
\begin{array}{l}
\phi^*+\delta _2+\delta _1+\Delta (\phi^*+\delta _2+\alpha _1\delta
_1)+\Delta (\phi^*+\delta _2+\alpha _2\delta _1+\Delta (\phi^*+\delta
_2+\alpha _1\delta _1)) \\ \phi^*+\delta _3+\delta _1+\Delta
(\phi^*+\delta _3+\beta _1\delta _1)+\Delta (\phi^*+\delta _3+\beta
_2\delta _1+\Delta (\phi^*+\delta _3+\beta _1\delta _1)) 
\end{array}
\right. $

where $\alpha _1$, $\alpha _2$,
$\beta _1$, $\beta _2$ are the unknown fractions 
of the driving $\delta _1$ the
oscillators have run at the moment they receive the 
pulses from the white neighboring sites.  At this
point we linearize $\Delta$ around $\phi^{*}$:

$\Delta(\phi^{*}+\delta _2+\alpha _1\delta _1)=
\underbrace{\Delta(\phi^{*})}_{\Delta^*} +(\delta _2+\alpha _1\delta
_1)\underbrace{\Delta^{\prime }(\phi^{*})} _{\Delta^{\prime *}}$

 $\Delta (\phi^{*}+\delta _2+\alpha _2\delta _1+\Delta (\phi
^*+\delta _2+\alpha_1\delta _1))= \\
\hspace*{4cm}=
\underbrace{\Delta (\phi^{*}+\Delta
^*)}_{\Delta^{**}} +(\delta _2+\alpha_2\delta_1+(\delta_2+ \alpha
_1\delta _1)\Delta ^{\prime *})\underbrace{\Delta ^{\prime }(\phi
^*+\Delta ^*)}_{\Delta ^{\prime **}}$

With this linearization, the previous structure can be written for
the black oscillators as
\begin{equation} \mbox{B}\left\{ \begin{array}{l} \phi^{*}+\Delta
^*+\Delta ^{**}+C\delta _2+A\delta _1 \\ \phi^{*}+\Delta^*+\Delta
^{**}+C\delta _3+B\delta _1 \end{array} \right.  \label{wbdos}
\end{equation}
where $A$, $B$, and $C$ have the following expressions:
\begin{eqnarray} A &=&1+\alpha _1\Delta ^{\prime *}(1+\Delta
^{\prime **})+\alpha _2\Delta ^{\prime **} \nonumber \\ B
&=&1+\beta _1\Delta ^{\prime *}(1+\Delta ^{\prime **})+\beta
_2\Delta ^{\prime **} \label{ABC} \\ C &=&\Delta ^{\prime *
}+1+\Delta ^{\prime **}+\Delta ^{\prime *}\Delta ^{\prime
**}=(1+\Delta ^{\prime *})(1+\Delta ^{\prime **}) \nonumber
\end{eqnarray}
Now, we drive the population until the first black oscillator
arrives to the threshold, that will be a driving $1-(\phi^{*}+\Delta
^*+\Delta ^{**}+C\delta _2+A\delta _1)$, equal to $\phi^{*}-A\delta
_1-B\delta _2$, due to the fixed point condition.  The new
configuration is thus

$\mbox{W}\left\{ \begin{array}{l}
\phi^{*}+(1-A)\delta _1-C\delta _2 \\ \phi^{*}-A\delta _1-C\delta _2
\end{array} \right.  \\ \mbox{B}\left\{ \begin{array}{l} 1
\\ 1-((A-B)\delta _1+C(\delta _2-\delta _3)) \end{array} \right.
$

These steps explain exactly half a cycle of the process.  
Then we can define a new set of perturbations as
\begin{eqnarray} \delta _1^{\prime } &=&-C(\delta _3-\delta
_2)-(B-A)\delta _1 \nonumber \\ \delta _2^{\prime } &=&\delta
_1-C\delta _2-A\delta _1 \label{newdeltas}\\ \delta _3^{\prime }
&=&-C\delta _2-A\delta _1 \nonumber \end{eqnarray}
Now we can compare with the initial configuration.
The white
oscillators are bounded between $\phi^{*}+\delta_3^{\prime}$ and
$\phi^{*}+\delta_2^{\prime}$, whereas the black ones are between
$1-\delta _1^{\prime}$ and $1$.
Then the transformation matrix
for the perturbations $\delta _1$, $\delta _2$, $\delta _3$ is
\begin{equation} \left( \begin{array}{ccc} A-B & C & -C \\ 1-A & -C &
0 \\ -A & -C & 0 \end{array} \right) \label{matrix} \end{equation}
The eigenvalues of (\ref{matrix}) give the information about the
linear stability of the fixed points.  They are:
\begin{equation} 
\lambda _1 =-C,
\nonumber \\ 
\hspace{3em}
\lambda _{2,3} =
\frac{A-B\pm\sqrt{(A-B)^2+4C}}{2}  \end{equation}
Now we use that $\Delta^{\prime}(\phi) > -1$ always, and then $C>0$.
The absolute value of the eigenvalues will be
\begin{equation} 
|\lambda _1| =C, \nonumber 
\hspace{3em} 
|\lambda _{2,3}| <
\frac{|A-B| + \sqrt{|A-B|^2+4C}}{2} 
\nonumber 
\end{equation}
Due to the fact that $0< \alpha_i, \beta_i <1$ (by definition),
we have, for a decreasing PRC and using Eq. (\ref{ABC}), that
$C<A,B<1$, that is, $|A-B|<1-C$. This yields
\begin{equation} 
|\lambda _1| =C <1,\nonumber  
\hspace{3em} 
|\lambda _{2,3}| <
\frac{1}{2}\left(1-C + \sqrt{(1-C)^2+4C} \right) = 1
\nonumber 
\end{equation}
The fact that the three eigenvalues are less than one, in absolute 
value, ensures the stability of the fixed point, when the PRC is
a decreasing function.

On the other hand, for $\Delta^{\prime}(\phi)>0$ and using again that
$0< \alpha_i, \beta_i <1$ and Eq. (\ref{ABC}), we have $1<A, B<C$,
which implies $|A-B|<C-1$, and then $|\lambda _1| =C>1$,
$|\lambda _{2,3}| < C$.
The fact of having at least one eigenvalue with absolute
value larger than one ensures that the fixed point is unstable
if the PRC is an increasing function.
\subsection{Generalization to other structures}
The analysis of the 'domino' structure in 1d follows the same
ideas developed for the chessboard structure if an excitatory 
coupling is assumed. The only difference
is that a given unit receives always a pulse from another unit
of the same color, and a further
pulse from a unit of different color. This fact introduces
new values for $A$, $B$, and $C$ in the transformation matrix
(\ref{matrix}). In particular, the value of $C$ is
given by
$C= (1+ \Delta^{\prime *})$.
The analysis of the transformation matrix of the perturbations
shows that all the eigenvalues $\lambda_i$ satisfy $ |\lambda_i| < 1$
provided $\Delta^{\prime }(\phi)<0$. 
As this transformation is iterated $n$ times
($n \rightarrow \infty$) the perturbations will tend to zero,
ensuring the linear stability of the structure.

A generalization of the results for the 1d lattice to the 2d is also
easy to carry out and can be applied to any structure shown in Fig. 3.
In these cases, the value of $C$ depends on the 
number of phase-locked neighbors ($a$) as
\begin{equation}
C= \underbrace{ (1+ \Delta^{\prime *})
(1+ \Delta^{ \prime **}) \ldots }_{
\hspace{2em} \mbox{$a$ times}}
\end{equation}
Again, following the same formalism it is easy to prove that the new 
structures satisfy the same stability criteria as those in one dimension.
\section{Conclusions}
In this paper we have studied the collective behavior of a lattice
model of
pulse-coupled nonlinear oscillators. Each single unit has its own
intrinsic dynamics and interacts instantaneously with its neighbors
when it reaches a threshold value. At this moment the oscillator is
reset. The response of an oscillator 
at the reset point plays a relevant
role when considering the dynamical properties of the model. The
existence of a refractory time is assumed throughout the paper.

The continuous dynamical evolution of each unit can be transformed
into a discrete one by looking at the system when a fixed
oscillator reaches the threshold. This picture is appropriate to 
describe the features of spatially periodic patterns which are
fixed points of the discrete dynamics,
and whose linear stability has been
analyzed. Our main result is that a monotonously decreasing 
(increasing) phase response curve makes these structures to be
linearly stable (unstable). However, a detailed analysis of their
basins of attraction is still missing and deserves future attention.

\begin{ack} The authors are indebted to
L.F. Abbott and A.V.M. Herz for very
fruitful discussions and to S.  Bottani for sending us a copy of
\cite{Bottani} prior to publication.  This work has been supported by
CICyT of the Spanish Government, grant \#PB94-0897.  \end{ack}

\end{document}